\documentclass[twocolumn,aps,prl,10pt]{revtex4}

\usepackage[english]{babel} 	
\usepackage{graphics} 
\usepackage[pdftex]{graphicx}
\usepackage{amssymb} 			
\usepackage{amsmath} 			
\usepackage{booktabs}
\usepackage{tabularx}
\usepackage{rotating}
\usepackage{color}
\usepackage{float}
\usepackage{url}	
\usepackage{bm}
\usepackage{epstopdf}

\usepackage{soul}

\DeclareMathAlphabet{\mathpzc}{OT1}{pzc}{m}{it}

\allowdisplaybreaks

\begin{document}


\title{Efimov effect in $D$ spatial dimensions in $AAB$ systems}

\author{D. S. Rosa$^{1}$, T. Frederico$^{2}$, G. Krein$^{1}$, M. T. Yamashita$^{1}$}

\affiliation{$^{1}$Instituto de F\'\i sica Te\'orica, Universidade
  Estadual Paulista, Rua Dr. Bento Teobaldo Ferraz, 271 - Bloco II, 01140-070 S\~ao Paulo, SP, Brazil}

\affiliation{$^{2}$Instituto Tecnol\'{o}gico de Aeron\'autica,
  12228-900, S\~ao Jos\'e dos Campos, SP, Brazil}

\begin{abstract}
The existence of the Efimov effect is drastically affected by the 
dimensionality of the space in which the system is embedded. The effective spatial 
dimension containing an atomic cloud can be continuously modified by compressing it 
in one or two directions. In the present article we determine  
for a general $AAB$ system formed by two identical bosons $A$ and a third particle $B$
in the two-body unitary limit, the dimensionsality $D$ for which the Efimov effect can 
exist for different values of the mass ratio $\mathpzc{A}=m_B/m_A$. In addition, we provide 
a prediction for the Efimov discrete scaling factor, ${\rm exp}\,(\pi/s)$, as a function 
of a wide range of values of $\mathpzc{A}$ and $D$, which can be tested in experiments 
that can be realized with currently available technology.
\end{abstract}

\maketitle

{\it Introduction.}  The Efimov effect~\cite{efimov} appears over an incredibly large variety 
of systems and spans a wide range of scales{\textemdash}atoms, molecules, exotic nuclei, 
condensed matter systems and even the DNA{\textemdash} turning into a quite rich research area 
in physics~\cite{Naidonreview2017}. Such a surprising phenomenon that can occur in a system of 
three particles with finite-range forces was discovered by V. Efimov in 1970 within 
the nuclear physics context, by showing that the system can have an infinite sequence of 
geometrically spaced energy levels when the pairs of the three particles have infinite 
two-body scattering length. Such a spectrum is closely related to the collapse of the three-body 
binding energy, discovered by L.H. Thomas in 1935~\cite{thomas}, by decreasing the range of the 
interaction with respect to its scattering length. 

The Thomas collapse and the Efimov effect are intimately related to the dimension where the system 
is inserted. Both Thomas and Efimov considered only a three dimensional environment ($D=3$). However, 
the flexibility of manipulating ultracold atomic traps brought the study of few-body systems 
to a new era: the Efimov effect was experimentally confirmed~\cite{kraemernature} after almost 40 
years since the original prediction; the use of the Feshbach resonance technique allows a fine 
tuning of the two-body interactions~\cite{kohlerrmp}; and the possibility to compress and expand 
the atomic cloud by changing the lasers and magnetic fields can create effectively two- and 
one-dimensional situations~\cite{pethick}.

Despite of the many advances in theory and experiment, there is an almost unexplored issue: 
the effect of a continuous changing of the spatial dimension
on the different observables. For integer dimensions, we know that the Efimov effect 
does not exist in $D=2$ or $D=1$.  More precisely, it has been demonstrated 
for identical bosons that the Efimov effect survives for $2.3 < D < 3.8$~\cite{nielsen}.  
However, the possibilities in engineering atomic traps with heteronuclear species 
raises the question on how the spatial dimensionality affects 
mass-imbalanced three-body systems. Such systems provide a more favorable situation 
for the experimental investigation of the Efimov effect~\cite{BaronPRL2009,BloomPRL2013,
PiresPRL2014,TungPRL2014,MaierPRL2015,ulmanis,WackerPRL2016,JohanNPHYS2017}.
Notwithstanding the fact that an atomic cloud compressed in one or two 
directions{\textemdash}when it acquires, respectively, a pancake or cigar shape{\textemdash}is
rigorously always a $D=3$ system, once the cloud size in one direction starts to become 
much smaller than the other, the system effectively starts to be affected by the dimensional 
reduction. This means that at some point, the excited three-body bound Efimov states  
start disappearing one-by-one, reflecting that the system is entering into a $D=2$ regime. 

In a broader context the continuous change in dimension affects the
transition from few to many body physics in cold atomic traps, as was discussed for Fermi gases 
in arbitrary dimensions~\cite{ValientePRA2012}, impurities forming polaritons 
within mass-imbalanced mixtures~\cite{BellottiNJP2015,SunPRL2017}, and now also with 
the feasibility of creating Efimov molecules from a resonantly interacting Bose 
gas~\cite{KlausPRF2017}. The change in dimension would also certainly affect systems 
formed by four mass-imbalanced atoms~\cite{ScmicklerPRA2017}. To the best of our knowledge, 
yet there is no experiment designed to investigate the response of a heteronuclear 
three-atom $AAB$ system to the change of dimensionality in a cold atomic trap close to a 
Feshbach resonance. 

In this article we consider mass imbalanced $AAB$ systems formed by two identical 
bosons ($A$) and a different particle ($B$) interacting by a zero-range pairwise 
interaction. We start from the Skorniakov and Ter-Martirosyan equations (STM)~\cite{stm} for 
the bound state of a system of three particles interacting pairwise through zero-range
interactions. We demonstrate that the STM coupled integral equations generalized to $D$ 
dimensions become scale invariant for large relative momenta $q$, so that the solutions 
are homogeneous functions of $q$ and have a log-periodic behavior for large $q$, 
a characteristic feature of the wave functions of the Efimov states. For three identical 
particles, our result for the critical interval of $D$ values where the Efimov state may 
be found, agrees with that reported in Ref.~\cite{nielsen}.

Our solutions, for a wide range of values of mass ratio $\mathpzc{A}=m_B/m_A$, 
provide predictions for the discrete Efimov scaling factor ${\rm exp}(\pi/s)$, 
which relates the three-body energies of consecutive excited states, 
$E_3^{(N)}/E_3^{(N+1)}={\rm exp}(2\pi/s)$, where $N=0,1,2...$ identifies the $N-$th 
Efimov state, with $N=0$ being the ground state.  Furthermore, the critical values of $D$ where 
these bound states appear are determined. We show how the asymmetric compression of the cloud 
can be associated with fractional dimensions.

{\it Methodology.} The Efimov effect is intimately related to the concept of universality. 
In the universal regime, the details of the short range part of the potential are not 
important for the low-energy properties of a system. This 
peculiar situation can be achieved once the two-body scattering 
length $a$ is much larger than the range $r_0$ of the potential. Then, a natural 
way to treat this regime is to choose a zero-range potential. 
Our calculation of the critical values of $D$, for which the Efimov effect 
ceases to exist, is based on two key points of the method proposed by 
Danilov~\cite{danilov} to find the  solution of STM  integral equation. 
The first key point is the dominance of the large momentum 
region for the solution of the STM equation due to the Thomas collapse. 
For large momentum, the finite values of the two- and three-body energies are 
irrelevant when compared to the kinetic energies and can then be set to zero.
The second key point is the emergence of invariance under a scale transformation
of the STM integral equations in the limit of large momentum. These two points are
enough for our purposes in the present work.

The set of coupled integral equations for the spectator functions, $\chi_A(q)$ and 
$\chi_B(q)$,  which generalizes the set of STM equations 
to  $D$ dimensions, reads (for simplicity, we take $\hbar=m_A=1$):
\begin{widetext}
\begin{eqnarray}
\label{chia}
\chi_{A}(q) &=& \tau_{AB}\left(E_3 -\frac{\mathpzc{A}+2}{2(\mathpzc{A}+1)}q^2\right) 
\int d^{D}k\left(\frac{\chi_{B}(k)}{E_3 - q^2 - \frac{\mathpzc{A}+1}{2\mathpzc{A}}k^2 
- \bold{k}\cdot\bold{q}}+
\frac{\chi_{A}(k)}{E_3  - \frac{\mathpzc{A}+1}{2\mathpzc{A}}(k^2+q^2) 
- \frac{1}{\mathpzc{A}}\bold{k}\cdot\bold{q}} \right) ,
\\ 
\label{chib}
\chi_{B}(q) &=& 2 \, \tau_{AA}\left(E_3 - \frac{\mathpzc{A}+2}{4\mathpzc{A}}q^2\right)  
\int d^{D}k \, \frac{\chi_{A}(k)}{E_3 - \frac{\mathpzc{A}+1}{2\mathpzc{A}}q^2 -k^2- 
\bold{k}\cdot\bold{q}}, 
\end{eqnarray}
where $E_3$ is the energy of the three-body system. The relative Jacobi momenta $\bold{q}$ and $\bold{k}$ 
are defined such that their origin is the center-of-mass of a given pair and point towards the remaining 
particle. The two-body transition amplitudes $\tau_{AB}$ and $\tau_{AA}$ are given by
\begin{eqnarray}
\label{tauab}
&&\tau^{-1}_{AB}\left(E_3 -\frac{\mathpzc{A}+2}{2(\mathpzc{A}+1)}q^2\right)  = 
\int d^D k \left(\frac{1}{- |E_2^{AB}|- \frac{{\mathpzc{A}}+1}{2{\mathpzc{A}}}k^2} -  
\frac{1}{E_3 - \frac{\mathpzc{A}+2}{2(\mathpzc{A}+1)} q^2 - \frac{{\mathpzc{A}}+1}{2{\mathpzc{A}}}k^2}
\right),\\
\label{tauaa}
&&\tau^{-1}_{AA}\left(E_3 - \frac{\mathpzc{A}+2}{4\mathpzc{A}}q^2\right) = 
\int d^D k \left(\frac{1}{- |E_2^{AA}|- k^2} - \frac{1}{E_3 - \frac{\mathpzc{A}+2}{4\mathpzc{A}} q^2 - k^2}\right),
\end{eqnarray}
where $E_2^{AB}$ and $E_2^{AA}$ are the two-body energies 
of the bound $AB$ and $AA$ systems, respectively. For large values of~$q$, relevant for exploring the 
Efimov effect, the integrals in Eqs.~(\ref{tauab}) and (\ref{tauaa}) are determined by the region 
of large values of $k$~\cite{danilov} and, therefore, the energies $E_3$, $E_2^{AA}$ and $E_2^{AB}$
can be set to zero in those integrals {\textemdash} this is the first point noted by Danilov. 
In this situation, one can obtain closed forms for the amplitudes $\tau_{AB}$ and $\tau_{AA}$:
\begin{eqnarray}
&&\tau^{-1}_{AB}\left(-\frac{\mathpzc{A}+2}{2(\mathpzc{A}+1)}q^2\right) = 
- q^{D-2} \left( \frac{\mathpzc{A}+2}{2(\mathpzc{A}+1)}  \right)^{{D/2-1}} 
\; \left(\frac{2\mathpzc{A}}{\mathpzc{A}+1}\right)^{{D}/{2}} \Gamma \left(D/2-1\right)
\Gamma\left(2-D/2\right) \frac{\pi^{{D}/{2}}}{\Gamma({D}/{2})}, \\
&&\tau^{-1}_{AA}\left(- \frac{\mathpzc{A}+2}{4\mathpzc{A}}q^2\right)  =  - q^{D-2} 
\left( \frac{\mathpzc{A}+2}{4\mathpzc{A}}  \right)^{{D/2-1}} 
\, 
\Gamma \left({D/2-1}\right) \Gamma\left({2-D/2}\right)
\frac{\pi^{{D}/{2}}}{\Gamma\left({D}/{2}\right)},
\end{eqnarray}
\end{widetext}
where $\Gamma(z)$ is the gamma function, defined for all complex numbers $z$ except for the 
non-positive integers. This restricts the validity of our results to the interval $2<D<4$.  
Their solutions are homogeneous functions, that is, the amplitudes $\chi_A(q)$ and $\chi_B(q)$ 
are given by
\begin{equation}
\chi_A(q) = C_A \, q^{r+is} \,\,\,{\rm and}\,\,\, \chi_B(q) = C_B \, q^{r+is},
\label{ansatz}
\end{equation}
where $r$ and $s$ are real numbers {\textemdash} this is the second point noted by Danilov. 
These solutions are the well-known log-periodic functions, 
associated with the infinitely many three-body bound states in the Efimov limit. 
Using Eq.~(\ref{ansatz}) in Eqs.~(\ref{chia}) and (\ref{chib}) leads to a complex homogeneous 
linear matrix equation for the coefficients $C_A$ and $C_B$. The parameters $r$ and $s$ 
are found by solving the corresponding characteristic equation. 
The real part of the 
exponent is given by the ansatz $r = 1 - D$ for all $D$, which removes any ultraviolet divergence.
For $D=3$ and ${\mathpzc{A}}=1$ one obtains $s=1.00623$, value that agrees with the well 
known result from Efimov~\cite{efimov}. Moreover, for $r=1-D$, 
the characteristic equation reads:
\begin{widetext}
\begin{equation} 
{\cal F}_D \left[\mathpzc{A} \, I_1(\mathpzc{A},s)  + 2  
\left( \frac{4\mathpzc{A}}{\mathpzc{A}+2} \right)^{{D/2-1}} 
{\cal F}_D \, I_2(\mathpzc{A},s) I_3(\mathpzc{A},s) \right]
= \left( \frac{\mathpzc{A}+2}{2(\mathpzc{A}+1)}  \right)^{{D/2-1}}  
\left(\frac{2\mathpzc{A}}{\mathpzc{A}+1}\right)^{{D}/{2}} ,
\label{eqfinal}
\end{equation}
\end{widetext}
where
\begin{equation}
{\cal F}_D=\frac{1}{\Gamma\left({D/2-1}\right) 
\Gamma\left({2-D/2}\right)},
\end{equation} 
and 
\begin{eqnarray}
\hspace{-0.5cm}
I_{1}(\mathpzc{A},s) \!&=&\! \int^{\infty}_0 \!dz\frac{z^{is}}{z} 
\log \left[ \frac{(z^{2}+1)(\mathpzc{A}+1) +2z}{(z^{2}+1)(\mathpzc{A}+1) -2z}\right], \\
\hspace{-0.5cm}
I_{2}(\mathpzc{A},s) \!&=&\! \int^{\infty}_0 \!dz\frac{z^{is}}{z} 
\log \left[ \frac{2\mathpzc{A}(z^{2}+z) + \mathpzc{A}+1}{2\mathpzc{A}(z^{2}-z)+\mathpzc{A}+1}\right], \\
\hspace{-0.5cm}
I_{3}(\mathpzc{A},s) \!&=&\! \int^{\infty}_0 \!dz\frac{z^{is}}{z} \log \left[ \frac{2\mathpzc{A}(1+z) 
+ (\mathpzc{A}+1)z^{2}}{2\mathpzc{A}(1-z) + (\mathpzc{A}+1)z^{2}}\right] ,  
\end{eqnarray}
which are the same integrals found in Ref. \cite{yamashitaPRA2013} for the $D=3$ problem. We note
that the result $r = 1 - D$ is exact. This can be proved by setting $r = 1 - D + \epsilon$ and
expanding the characteristic equation in a power series in $\epsilon$: for given 
values of $D$ and $\mathpzc{A}$, it can be verified analytically that the only possible 
solution occurs for $\epsilon = 0$.

{\it Results.} In cold-atom traps when a three-atom bound state crosses the continuum threshold 
the atoms can recombine forming a deeply bound two-atom molecule plus 
an atom. The recoil energy of the 
atom-molecule system is much larger than the depth of the ultracold trap in such a way the three 
atoms are lost. The three-atom recombination peaks appear at two-body scattering lengths, 
$a_-^{(N)}$, separated by multiplicative factors of ${\rm exp}(\pi/s)$. Deviations from 
the $D=3$ limit, excluding range corrections, are associated with the response of the 
three-body system to the dimension changes between $D=2$ and $D=3$. 
Note that in heteronuclear systems two scattering lengths can be distinguished: one for 
the $AA$ subsystem, and another one for the $AB$ subsystem. The present solution given by 
Eq. (\ref{ansatz}) corresponds to the limit of both scattering lengths 
tending to infinity.

\begin{figure}[htb!]
\includegraphics[width=9cm]{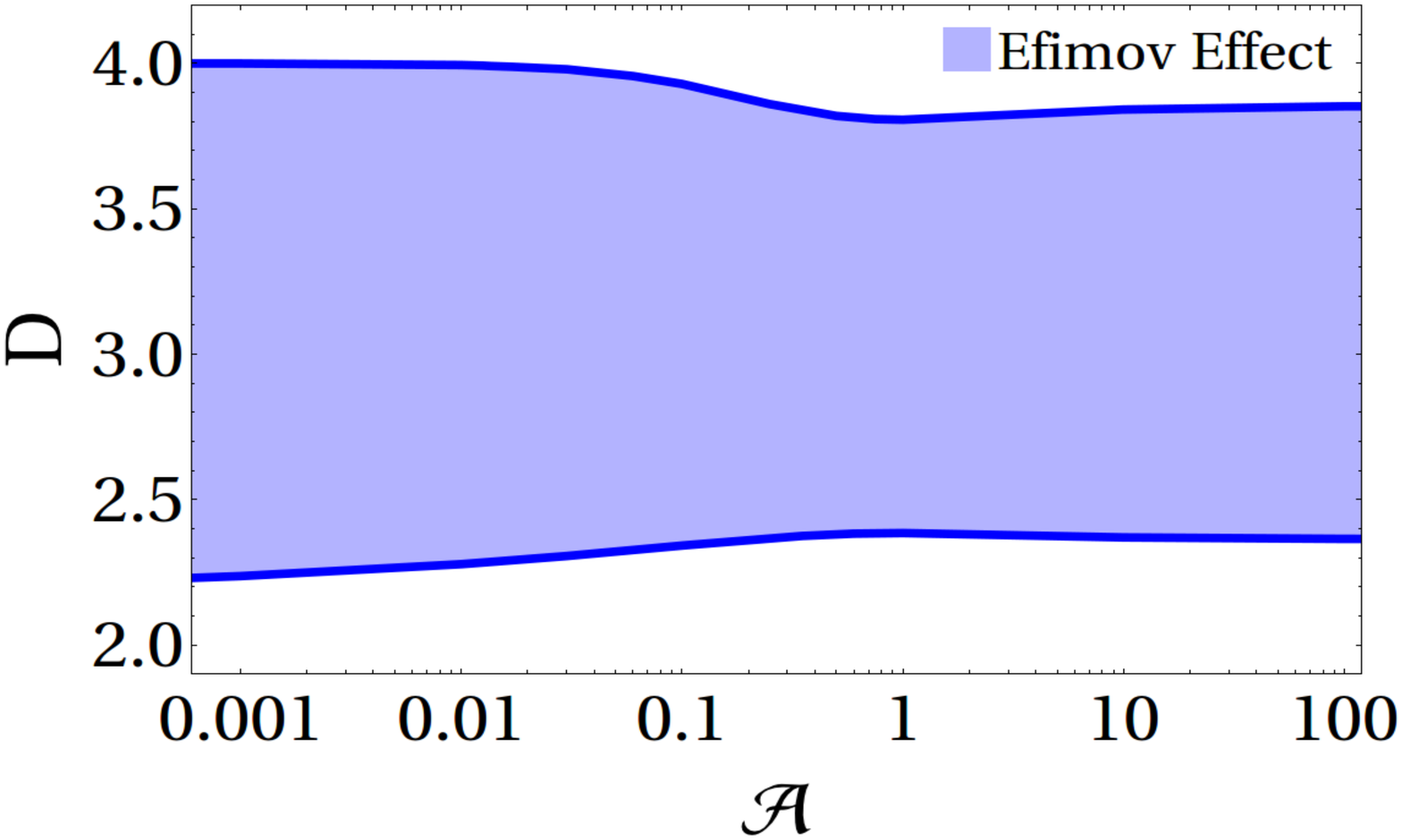}
\caption{Regions (in blue) where there is a real solution for the scaling factor, $s$, solution
to the Eq.~(\ref{eqfinal}); outside this ``dimensional band'', the Efimov effect does not exist. 
For $\mathpzc{A}=1$ we reproduce exactly the result in Ref.~\cite{nielsen}, where the 
dimensional limits are given by $2.3<D<3.8$. } 
\label{boundaries}
\end{figure}

Once $\mathpzc{A}$ is fixed, the imaginary part $s$ of the exponent of $q$ in Eq.~(\ref{ansatz}) 
is the solution of Eq. (\ref{eqfinal}). The boundaries of the region of values of $D$ for which
the Efimov effect survives are determined by the existence of nonzero values of~$s$; close to the 
threshold, the Efimov effect disappears as $s \to 0$ and the energy gap between 
levels tends to infinity. The boundaries are shown in Fig.~\ref{boundaries}.

In experiments, it is possible to change the confining potential 
in order to squeeze one or two directions of the trap transforming the 
cloud in a quasi-($D=2$) or quasi-($D=1$) environment, respectively. Rigorously, as mentioned,
all these systems are in $D=3$; however, the three-body system embedded in the atomic cloud 
feels an effective dimension when compressed{\textemdash}as shown in previous 
works~\cite{compactPRA,levinsen}{\textemdash} that makes the most excited Efimov 
states disappear one-by-one until reaching the expected number of bound 
states in $D=2$.

The physical reason behind the disappearance of the Efimov states close to 
the critical dimension can be easily understood considering the Born-Oppenheimer (BO)
approximation, valid in the situation $m_A \gg m_B$. In the BO
approximation, an effective potential coming from the exchange of the
light particle between the two heavy ones can be extracted. 
The form of
this potential is well known in $D=3$, given by Ref.~\cite{fonseca}, $- (s + 1/4)/R^2$,  
where $R$ is the separation distance between the heavy particles,
and $s$ is the imaginary part of the exponent of $q$ in Eq.~(\ref{ansatz}). 
The Efimov effect is due to the ``fall to the center'' for $s>-1/4$. 
For heteronuclear systems in $D$ dimensions the effective potential is still proportional 
to $-1/R^2$, but the strength is now more complicated depending on $D$ and 
$\mathpzc{A}$. For a given mass ratio, at the critical dimensions on either side of 
$D=3$, i.e. $D > 3$ and $D < 3$, the Efimov effect disappears precisely at the critical 
strength $-(D-2)^2/4$~\cite{wip}, reproducing the result for $D=3$~\cite{fonseca}, where the 
fall to the center stops.

\begin{figure}[htb!]
\centering
\includegraphics[width=8cm]{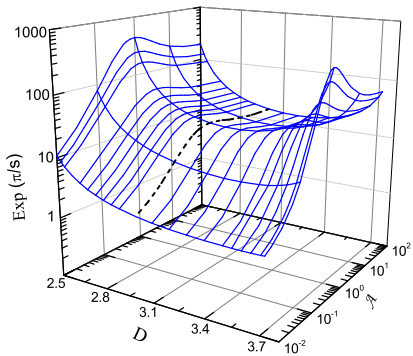}
\caption{Discrete scaling factor as a function of the mass ratio $\mathpzc{A}=m_B/m_A$, 
and dimension $D$. The black dashed line shows the well-known situation of $D=3$.} 
\label{efimovfactor}
\end{figure}

Figure~\ref{efimovfactor} shows the value of the discrete scaling factor 
${\rm exp}\,(\pi/s)$ for a wide range of the mass ratio $\mathpzc{A}$ and 
of the band of $D$ values that includes the only allowed integer dimension ($D=3$) for 
which the Efimov effect exists. The black dashed line indicates the 
well-known result for $D=3$~\cite{braaten}. The most symmetrical case, where $\mathpzc{A}=1$, 
presents the worst situation to observe consecutive Efimov excited states as for 
any $D$ the scaling factor presents a maximum for this mass ratio. 

\begin{figure}[thb!]
\includegraphics[width=7cm]{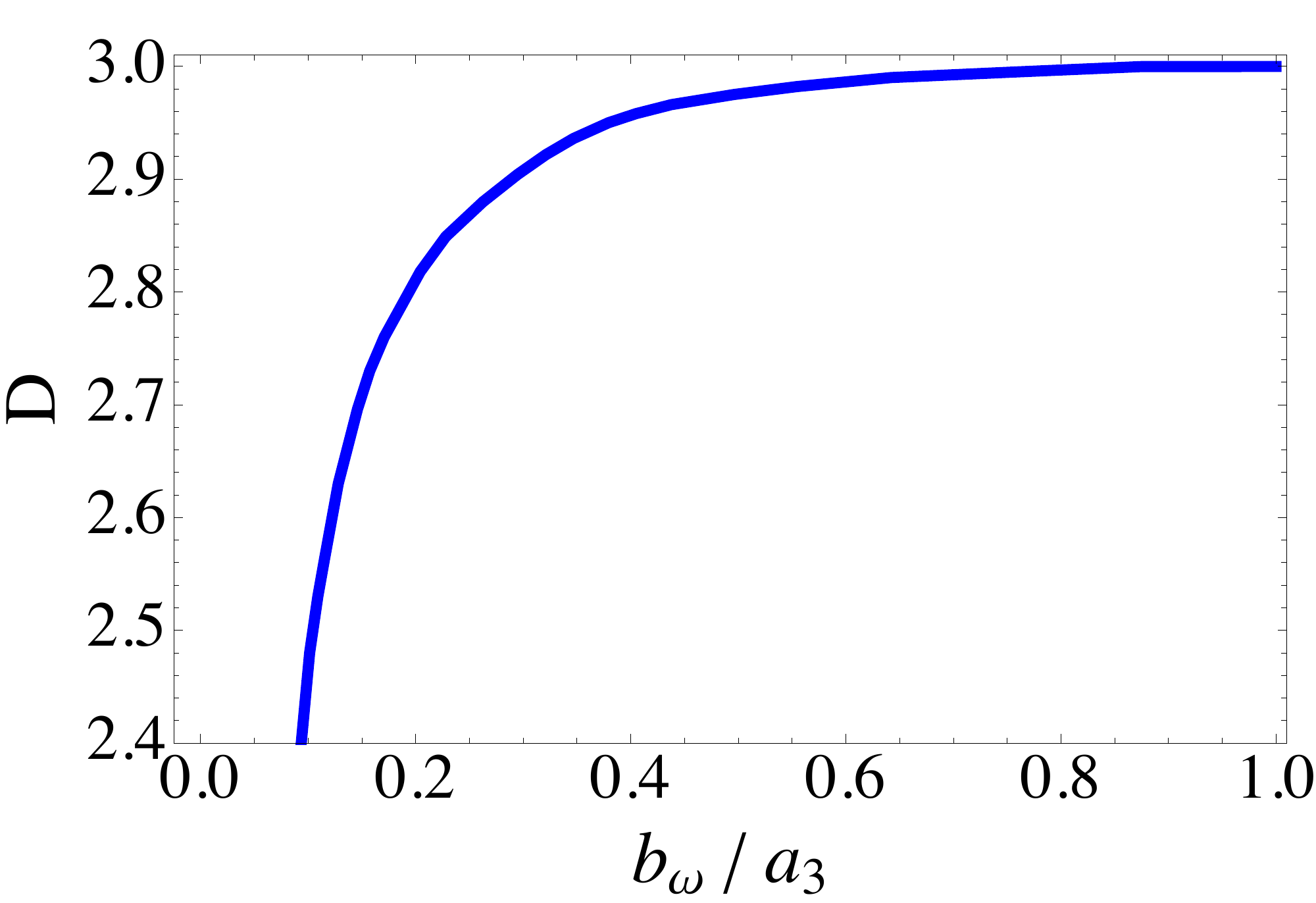}
\caption{Effective dimension $D$ for $\mathpzc{A}=6/133$ as a function of 
$b_\omega/a_3$, where $a_3$ is the $AB$ scattering length for $D=3$.}
\label{dbw}
\end{figure}

{\it Experimental Possibilities.} A connection between our calculations and finite 
energy situations can be made as follows. In a system confined by a squeezed 
harmonic trap with an oscillator length in the squeezed direction given 
by $b_\omega$, the two- and three-body energies, respectively, $E^{AB}_2$ 
and $E_3$ can be calculated by solving the Faddeev equations in 
momentum space with a compactified dimension as detailed in Ref.~\cite{john}. 
From Fig.~2 of Ref.~\cite{john} one can extract $s = s(b_\omega)$ that gives 
a relationship between $b_\omega$ and $D$ through the Eq.~(\ref{eqfinal}). This 
relationship should be adequate for an experimental assessment of our prediction. 
To illustrate the principle, we consider the realistic case of a $^6$Li-$^{133}$Cs 
mixture, for which $\mathpzc{A}=6/133$. The results for $D$ are shown in  Fig.~\ref{dbw} as a function 
of the ratio $b_\omega/a_3$, where $a_3$ is the $AB$ scattering length for $D=3$, for the situation 
in which $E^{AB}_2$ is kept fixed to its value for $D=3$ (dashed curves in Fig.~2 of Ref.~\cite{john}).

It is important to note that, for finite two and three-body energies, when $b_\omega$ is  
much larger than the size of the three-body system, all states experience the same effective 
dimension. Therefore, the discrete scaling factor can be used to infer the effective 
dimension during the squeezing process along the region where these trimer states exist.
When $b_\omega$ becomes of the order of the size of the state, this is not possible anymore. 
In this situation, the three-body bound states turn into a virtual or resonant state, 
depending whether the two-body state is bound or virtual, respectively. Likewise, 
the features of our predictions for the scaling factor as a function of 
$D$ and $\mathpzc{A}$ shown in Fig.~\ref{efimovfactor} are not expected to be washed out 
away from the unitary limit and should be tested experimentally by measuring the ratio of 
the scattering lengths at the peaks of the atomic loss rate, associated with the appearance 
of three-body bound states.

Experimentally, it is not possible to reach the exact unitary limit, 
i.e. exact zero two-body binding energy. The discrete scaling symmetry away from
the unitary limit has been well established for $D=3$ in homonuclear systems. For 
finite scattering lengths deviations from the exact unitary symmetry have been 
observed in the $^{133}$Cs-$^{133}$Cs-$^6$Li system for $a_{BB}>0$ and $a_{AB}<0$~\cite{ulmanis}. 
In the same way, deviations from our predictions are also expected for non-integer 
dimensions. In an experiment it is also not possible to access very highly 
excited states. However, the measurement of consecutive three-body bound states {\textemdash} 
revealed by the appearance of peaks in the atomic loss rate {\textemdash} can be observed 
by choosing systems with sufficiently large mass asymmetry, as in 
this case the gap between consecutive energy levels is reduced. Corrections due to 
the effect of finite two-body binding energy affects considerably only the scale 
factor extracted from the ground state, as shown in Ref.~\cite{mathias}. If higher 
excited states are considered to calculate $s$, the scale factor will be practically 
given by the values predicted in this paper.

{\it Conclusion.} In this article we have studied the dimensional limits for the occurrence of 
the Efimov effect for a general mass ratio $\mathpzc{A}=m_B/m_A$. Our calculation 
is performed in the unitary limit and reproduces the well-known result for 
$\mathpzc{A}=1$, where the limit is given by $2.3<D<3.8$~\cite{nielsen}, and generalizes it 
for a wide range of $\mathpzc{A}$ and $D$. We also predict the numerical values of the discrete 
scaling factor ${\rm exp}\,(\pi/s)$ as a function of $\mathpzc{A}$ and $D$, which can be 
tested in experiments with currently available technology. 

\paragraph*{Acknowledgments.} This work was partly supported by funds provided by 
the Brazilian agencies Funda\c{c}\~{a}o de Amparo \`{a} Pesquisa do Estado de 
S\~{a}o Paulo - FAPESP grants no. 2016/01816-2(MTY), 2013/01907-0(GK) and 2017/05660-0(TF), 
Conselho Nacional de Desenvolvimento Cient\'{i}fico e Tecnol\'{o}gico - 
CNPq grants no. 305894/2009(GK), 302075/2016-0(MTY), 308486/2015-3 (TF),
Coordena\c{c}\~{a}o de Aperfei\c{c}oamento de Pessoal de N\'{i}vel Superior - 
CAPES grant no. 88881.030363/2013-01(MTY). 
This work is a part of the project INCT-FNA Proc. No. 464898/2014-5. 
We thank John Sandoval for providing the data for Fig.~3.


\begin{thebibliography}{28}

\bibitem{efimov} V. Efimov, Phys. Lett. B {\bf 33}, 563 (1970).

\bibitem{Naidonreview2017} P. Naidon and S. Endo, Rep. Prog. Phys. {\bf 80}, 
056001 (2017).

\bibitem{thomas} L. H. Thomas, Phys. Rev. {\bf 47}, 903 (1935).

\bibitem{kraemernature} T. Kraemer, M. Mark, P. Waldburger, J. G. Danzl, C. Chin, 
B. Engeser, A. D. Lange, K. Pilch, A. Jaakkola, H.-C. Nägerl and R. Grimm, 
Nature {\bf 440}, 315 (2006).

\bibitem{kohlerrmp} T. K\"ohler, K. Goral and P. S. Julienne, Rev. Mod. Phys. 
{\bf 78}, 1311 (2006).

\bibitem{pethick} C. J. Pethick and H. Smith, Bose-Einstein Condensation 
in Dilute Gases (Cambridge, 2008).

\bibitem{nielsen} E. Nielsen, D. V. Fedorov, A. S. Jensen and E. Garrido, 
Phys. Rep. {\bf 347}, 373 (2001).

\bibitem{BaronPRL2009} G. Barontini et al., Phys. Rev. Lett. {\bf 103}, 043201 (2009).

\bibitem{BloomPRL2013} R. S. Bloom et al., Phys. Rev. Lett. {\bf 111}, 105301 (2013).
 
\bibitem{PiresPRL2014} R. Pires et al., Phys. Rev. Lett. {\bf 112}, 250404 (2014).
 
\bibitem{TungPRL2014} S. K. Tung et al., Phys. Rev. Lett. {\bf 113}, 240402 (2014).
 
\bibitem{MaierPRL2015}  R. A. W. Maier et al., Phys. Rev. Lett. {\bf 115}, 043201
(2015).

\bibitem{ulmanis} J. Ulmanis, S. H\"afner, R. Pires, E. D. Kuhnle, Yujun Wang, 
Chris H. Greene and M. Weidem\"uller, Phys. Rev. Lett. {\bf 117}, 153201 (2016). 

\bibitem{WackerPRL2016} L. J. Wacker et al., Phys. Rev. Lett. {\bf 117}, 163201 (2016).

\bibitem{JohanNPHYS2017} J. Johansen, B. J. DeSalvo, K. Patel and C. Chin, Nature Phys. 
{\bf 13}, 731 (2017).

\bibitem{ValientePRA2012} M. Valiente, , N. T. Zinner and K. Molmer, Phys. Rev. A 
{\bf 86}, 043616 (2012).

\bibitem{BellottiNJP2015} F. F. Bellotti, T. Frederico, M. T. Yamashita, D. V. Fedorov, 
A. S. Jensen and N. T. Zinner, New J. Phys. {\bf 18}, 043023 (2016).

\bibitem{SunPRL2017} M. Sun, H. Zhai and X. Cui, Phys. Rev. Lett. {\bf 119}, 013401 (2017).

\bibitem{KlausPRF2017} C. E. Klauss, X. Xie, C. Lopez-Abadia, J. P. D.Incao, Z. Hadzibabic, 
D. S. Jin and E. A. Cornell, Phys. Rev. Lett. {\bf 119}, 143401 (2017).

\bibitem{ScmicklerPRA2017} C. H. Schmickler, H.-W. Hammer and E. Hiyama, 
Phys. Rev. A {\bf 95}, 052710 (2017).

\bibitem{stm} G.V. Skornyakov and K.A. Ter-Martirosyan, Zh. Eksp. Teor. Fiz. {\bf 31}, 
775 (1957).

\bibitem{danilov} G. S. Danilov, Sov. Phys. JETP {\bf 13}, 349 (1961).

\bibitem{yamashitaPRA2013} M. T. Yamashita, F. F. Bellotti, T. Frederico, 
D. V. Fedorov, A. S. Jensen and N. T. Zinner, Phys. Rev. A {\bf 87}, 062702 (2013).

\bibitem{compactPRA} M. T. Yamashita, F. F. Bellotti, T. Frederico, D. V. Fedorov, 
A. S. Jensen and N. T. Zinner, J. Phys. B {\bf 48}, 025302 (2015).

\bibitem{levinsen} J. Levinsen, P. Massignan and M. M. Parish, Phys. Rev. X 
{\bf 4}, 031020 (2014).

\bibitem{braaten} E. Braaten and H.-W. Hammer, Phys. Rep. {\bf 428}, 259 (2006).

\bibitem{fonseca} A. C. Fonseca, E. F. Redish and P. E. Shanley, Nucl. Phys. {\bf A320}, 273 (1979).

\bibitem{wip} D. S. Rosa, T. Frederico, G. Krein and M. T. Yamashita, work in progress.

\bibitem{john}  J. H. Sandoval et al., J. Phys. B: At. Mol. Opt. Phys. {\bf 51}, 065004 (2018).  

\bibitem{mathias} S. H\"afner, J. Ulmanis, E. D. Kuhnle, Y. Wang, C. H. Greene,
and M. Weidem\"uller, Phys. Rev. A {\bf 95}, 062708 (2017).

\end{thebibliography}
\end{document}